\documentclass[prb,aps, twocolumn, amsmath,amssymb,superscriptaddress]{revtex4}
\usepackage{braket}
\usepackage{adjustbox}
\usepackage{float}
\usepackage{amsmath}
\usepackage{amssymb}
\usepackage{amsfonts}
\usepackage{euscript}
\usepackage{enumerate}
\usepackage{hhline}
\usepackage{pslatex}
\usepackage{tabularx}
\usepackage[usenames,dvipsnames]{xcolor}
\usepackage{threeparttable}
\usepackage{graphicx}
\usepackage{dcolumn}
\usepackage{bm}
\usepackage[sort&compress]{natbib}

\makeatletter

\renewcommand{\@biblabel}[1]{#1. }
\renewcommand{\@dotsep}{500}
\renewcommand{\@pnumwidth}{0em}
\renewcommand{\l@figure}[2]{
\@dottedtocline{1}{1.5em}{2em}{Figure #1}{}\vspace{15pt}}

\begin{document}

\title{A solid-state entangled photon pair source with high brightness and indistinguishability}

\author{Jin Liu}
\thanks{These authors contributed equally}
\affiliation{State Key Laboratory of Optoelectronic Materials and Technologies, School of Physics, Sun Yat-sen University, Guangzhou 510275, China}

\author{Rongbin Su}
\thanks{These authors contributed equally}
\affiliation{State Key Laboratory of Optoelectronic Materials and Technologies, School of Physics, Sun Yat-sen University, Guangzhou 510275, China}

\author{Yuming Wei}
\affiliation{State Key Laboratory of Optoelectronic Materials and Technologies, School of Physics, Sun Yat-sen University, Guangzhou 510275, China}

\author{Beimeng Yao}
\affiliation{State Key Laboratory of Optoelectronic Materials and Technologies, School of Physics, Sun Yat-sen University, Guangzhou 510275, China}

\author{Saimon Filipe Covre da Silva}
\affiliation{Institute of Semiconductor and Solid State Physics, Johannes Kepler University, Altenbergerstrae 69, Linz 4040, Austria}

\author{Ying Yu}
\affiliation{State Key Laboratory of Optoelectronic Materials and Technologies, School of Electronics and Information Technology, Sun Yat-sen University, Guangzhou 510275, China}

\author{Jake Iles-Smith}
\affiliation{{School of Physics and Astronomy, The University of Sheffield, Sheffield, S10 2TN, United Kingdom}}

\author{Kartik Srinivasan}
\affiliation{Center for Nanoscale Science and Technology, National Institute of Standards and Technology, Gaithersburg, MD 20899, USA}

\author{Armando Rastelli}
\thanks{Correspondence to Armando.Rastelli@jku.at, lijt3@mail.sysu.edu.cn, wangxueh@mail.sysu.edu.cn}
\affiliation{Institute of Semiconductor and Solid State Physics, Johannes Kepler University, Altenbergerstrae 69, Linz 4040, Austria}

\author{Juntao Li}
\thanks{Correspondence to Armando.Rastelli@jku.at, lijt3@mail.sysu.edu.cn, wangxueh@mail.sysu.edu.cn}
\affiliation{State Key Laboratory of Optoelectronic Materials and Technologies, School of Physics, Sun Yat-sen University, Guangzhou 510275, China}

\author{Xuehua Wang}
\thanks{Correspondence to Armando.Rastelli@jku.at, lijt3@mail.sysu.edu.cn, wangxueh@mail.sysu.edu.cn}
\affiliation{State Key Laboratory of Optoelectronic Materials and Technologies, School of Physics, Sun Yat-sen University, Guangzhou 510275, China}

\date{\today}

\begin{abstract}
\noindent \textbf{The generation of high-quality entangled photon pairs has been being a long-sought goal in modern quantum communication and computation. To date, the most widely-used entangled photon pairs are generated from spontaneous parametric downconversion, a process that is intrinsically probabilistic and thus relegated to a regime of low pair-generation rates. In contrast, semiconductor quantum dots can generate triggered entangled photon pairs via a cascaded radiative decay process, and do not suffer from any fundamental trade-off between source brightness and multi-pair generation. However, a source featuring simultaneously high photon-extraction efficiency, high-degree of entanglement fidelity and photon indistinguishability has not yet been reported. Here, we present an entangled photon pair source with high brightness and indistinguishability by deterministically embedding GaAs quantum dots in broadband photonic nanostructures that enable Purcell-enhanced emission. Our source produces entangled photon pairs with a record pair collection probability of up to 0.65(4) (single-photon extraction efficiency of 0.85(3)), entanglement fidelity of 0.88(2), and indistinguishabilities of 0.901(3) and 0.903(3), which immediately creates opportunities for advancing quantum photonic technologies.}
	

\end{abstract}

 \maketitle

\maketitle
Quantum entanglement is one of the most intriguing properties in quantum physics\cite{EPR}, in which the quantum state of a many-particle system cannot be written as a product of the single-particle wave functions, no matter how far they are separated from each other. Entangled photon pairs, which are immune from decoherence and are easy to manipulate and detect, have played an essential role in the epic triumph of quantum physics over local causality through optical tests of Bell's inequalities\cite{Giustina2015,Shalm2015}. In the modern quantum technology era, entangled photon pairs serve as a key element in many quantum photonic information processing protocols\cite{Bouwmeester_book,Kimble2008}, e.g., the quantum repeater\cite{Simon2007} and device-independent quantum key distribution\cite{Acin2007}. To date, spontaneous parametric down conversion (SPDC)\cite{Kwiat1995,Geneva2005,Wang2016} is the most widely used "working horse" for generating entangled-photon pairs with high degree of entanglement fidelity and photon indistingshability\cite{Pan2012}. However, the Poissonian statistics of such sources intrinsically limits their brightness to an operation rate that is typically $<$~0.1\cite{Pan2012} (the average photon pair generation probability per pulse), thus imposing a great challenge in advancing efficiency-demanding photonic quantum technologies.

Alternatively, epitaxial semiconductor quantum dots (QDs) have been successfully demonstrated as a potentially scalable technology for triggered sources of entangled photon pairs via the biexciton (XX)~-~exciton (X) cascaded radiative processes\cite{Benson2000,Stevenson2006,Akopian2006,Muller2009,Michler2014,Pelucchi2016}. Their small footprint and compatibility with semiconductor technology make them particular appealing for on-chip integration\cite{Orieux2017}. However, a multitude of challenges have to be overcome to be able to realize optimal semiconductor sources of entangled photon pairs. First, the fine structure splitting (FSS) of the neutral exciton state, a result of the electron-hole exchange interaction in asymmetric QDs, reveals the radiative decay path information and consequently significantly reduces the time-averaged entanglement fidelity. This issue has been very recently alleviated by developing QDs with highly symmetric shapes, either by InGaAs growth in inverted pyramids\cite{Pelucchi2016} or by optimized droplet-etching \cite{Huo2013}. Furthermore, it has been argued that the comparatively short radiative lifetime of X, the small nuclear spin number of Ga (3/2 compare to 9/2 of In), and the use of two-photon excitation effectively suppress the spin-flip and carrier recapture processes, thus facilitating the achievement of a high-degree of entanglement fidelity and indistinguishability\cite{Keil2017,Huber2017}.

\begin{figure*}[ht!]
\begin{center}
\includegraphics[width=0.7\linewidth]{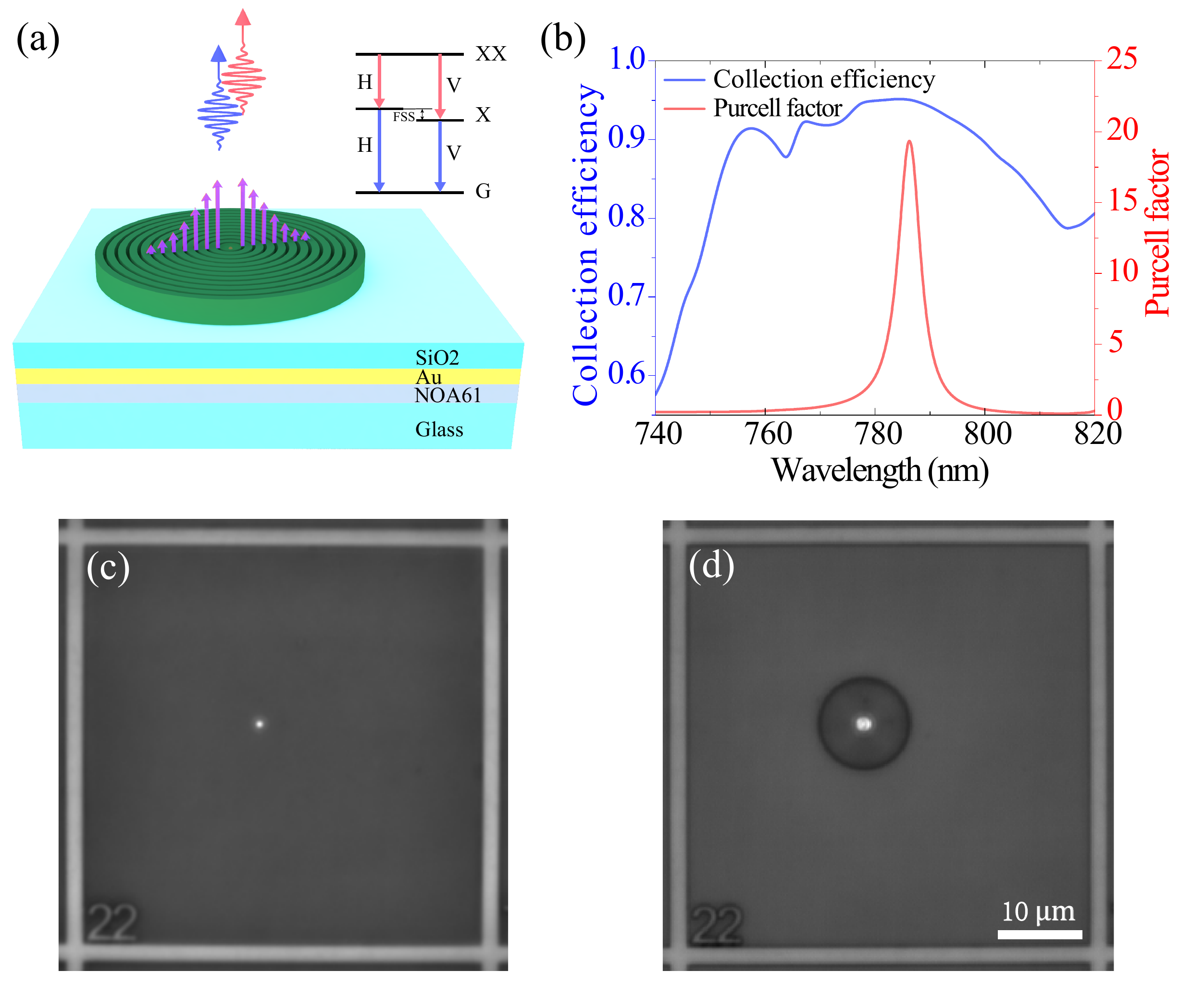}
\caption{Circular Bragg resonator on highly-efficient broadband reflector (CBR-HBR) for entangled-photon pair generation. Realization and calculated performance of the CBR-HBR are presented. (a) An illustration of a CBR-HBR with a single QD emitting entangled-photon pairs. The inset shows the XX-X cascaded radiative process for generating polarization-entangled photon pairs, in which the value of the fine structure splitting (FSS) plays an important role in determining the achievable entanglement fidelity without time-filtering. (b) Simulated Purcell factor (red) and collection efficiency (blue) of the CBR-HBR as a function of wavelength. The collection efficiency is based on a $40^{\circ}$ azimuth angle, corresponding to a numerical aperture (NA) $=$ 0.65. (c) and (d) are fluorescence images of the same QD before and after the fabrication of the CBR-HBR. (c) and (d) share the same scale bar.}
\label{fig:Fig1}
\end{center}
\end{figure*}

Second, low photon extraction efficiency, a result of the high refractive index of the semiconductor material surrounding the QDs, has long been recognized as a hurdle for quantum light sources based on QDs. Typically, only $<$1\% of the photons emitted by QDs in bulk material can be collected by a free-space lens or objective. Photonic nanostructures, e.g, cavities\cite{Ding2016,Somaschi2016,He2017}, waveguides\cite{Claudon2010,Reimer2012,Laucht2012,Arcari2014}, microlenses\cite{Gschrey2015} and circular Bragg gratings\cite{Davanço2011,Sapienza2015}, exhibit excellent performance in funneling the single-photons emitted by QDs into free-space or optical fibers, but directly implementing these nanostructures for entangled-photon pair generation is not straightforward. The state-of-the-art QD entangled photon pair sources are based on micro-pillar "molecules"\cite{Dousse2010}, photonic nanowires\cite{Jons2017} and optical antennas\cite{Chen2018}, in which each single-photon in the pair efficiently couples into a dedicatedly designed photonic channels, resulting in bright-entangled photon pairs with a high degree of entanglement fidelity. Nevertheless, the overall performance of these entangled pair sources must be significantly improved for most applications, in terms of simultaneously achieving high brightness, entanglement fidelity, and photon indistinguishability.

Here we take a further step towards entangled photon pair sources with high brightness and indistinguishability by combining GaAs QDs with new broadband photonic nanostructures, i.e., circular Bragg resonators on highly-efficient broadband reflectors (CBR-HBR). Using a wide-field QD positioning technique\cite{Sapienza2015,He2017,Liu2017_RSI}, we deterministically fabricate CBR-HBRs in which single GaAs QDs (see S.I.~I) are precisely located at the optimal position (the center of the cavity) for high-performance entangled photon pair generation. A single-photon collection efficiency of up to 0.85(3) for both X and XX is achieved, resulting in a record photon pair collection probability of 0.65(4) per excitation pulse. A high degree of single-photon purity of 99.8(1)\%, entanglement fidelity of 0.88(2), and indistinguishabilities of 0.901(3) and 0.903(3) are also simultaneously obtained.
\begin{figure*}[ht!]
\begin{center}
\includegraphics[width=0.7\linewidth]{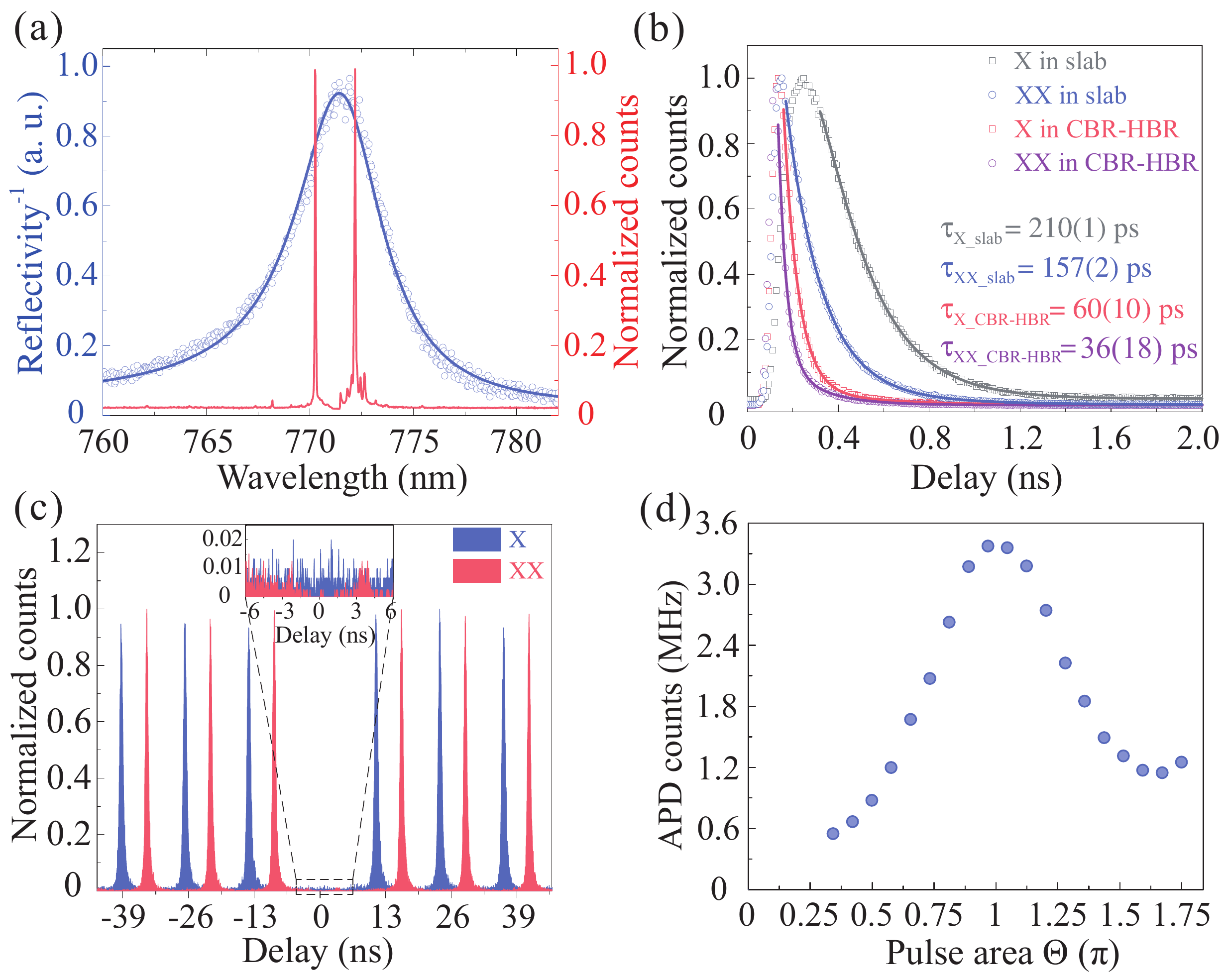}
\caption{\textbf{Basic characterization of the QD-CBR-HBR device.} (a) PL spectrum of a QD in the CBR-HBR under two-photon resonant excitation (right y axis, indicated in red) and the cavity mode measured from white light reflection (left y axis, indicated in blue). The excitation power is chosen to maximize the intensity of the XX emission ("$\pi$ pulse" conditions),  X and XX are equally populated and resonant with the cavity mode of the CBR-HBR. (b) PL lifetime of X and XX in bulk and in the CBR-HBR, showing pronounced Purcell enhancement for both X and XX. (c) Photon auto-correlation measured under "$\pi$ pulse" two-photon resonant excitation, using a Hanbury-Brown and Twiss interferometer. The second-order correlation $g^{(2)}(0)=0.001\pm0.001$ for X and $g^{(2)}(0)=0.007\pm0.001$ for XX are calculated from the integrated area in the zero delay peak divided by the mean of the peaks away from zero-delay, and the uncertainty is a one standard deviation value. (d) Detected count rates of the X photons as a function of square root of the excitation power. The blue curve is a guide to the eyes.}
\label{fig:Fig2}
\end{center}
\end{figure*}

\section*{Design and fabrication of devices}
In order to realize bright entangled photon pairs with Purcell-enhanced emission rates, we have developed a new nanostructure, i.e., CBR-HBR, with a few significant advantages respective to our previous work\cite{Sapienza2015}. Most notably, the implementation of the HBR strucuture effectively suppresses the downwards photon leakage and therefore significantly improves the collection efficiency over a broadband, see the details in the S.I.~II. Our CBR-HBR consists of a circular AlGaAs disk surrounded by a set of concentric AlGaAs rings, sitting on a $\mathrm{SiO_{2}}$ layer with a gold back reflector, as schematically shown in Fig.~1(a). The cavity resonance can be accurately engineered by varying the diameter of the central AlGaAs disk. Meanwhile, the in-plane emission is directed upwards by the concentric rings that meet the second-order Bragg conditions. By carefully designing the thickness of the $\mathrm{SiO_{2}}$ insulator layer, all the photons leaking into the substrate can be effectively reflected from the broadband gold mirror and recaptured by the CBR (See more details in S.I.~III). In such a situation, very high collection efficiencies can be obtained in a broadband manner. For QDs located in the center of the CBR, the simulated collection efficiency at the first lens and the Purcell factor as a function of the operation wavelength are plotted in Fig.~1(b). Collection efficiencies above 90~\% can be theoretically achieved in a bandwidth of $\approx$33~nm, and Purcell factors above 2 can be obtained for a bandwidth of $\approx$13~nm, which is 6.5 times the X-XX separation ($\approx$2~nm).

We have developed a membrane transfer technique to realize the $\mathrm{AlGaAs/SiO_{2}/Au}$ material platform from which the CBR-HBRs are fabricated, with the details provided in the S.I.~IV. We note that the presented photonic design is fully compatible with state-of-the-art piezoelectric-based tuning methods\cite{Yan2016,Trotta2016,Huber2018}, which enable the elimination of the FSS and the tuning of photon energy because of the flexible choice of substrate (here quartz) and flat morphology, which allows efficient strain transfer. By taking advantage of our recently developed QD positioning technique, we are able to identify individual QDs and extract their spatial positions with respect to alignment marks with an uncertainty of $\approx$10~nm\cite{Sapienza2015,Liu2017_RSI}, see Fig.~1(c). The CBR-HBR is then deterministically fabricated around the target QD. Figure~1(d) shows the fluorescence image of our device after the CBR-HBR fabrication, in which the targeted single QD in Fig.~1(c) is accurately located in the center of the fabricated CBR-HBR.

\section{Single-photon emission and brightness assessment}
Figure~2(a) presents the photoluminescence (PL) and white light reflectivity (1/R is shown, with R the reflectivity spectrum) of our device at 3.2 K, see the optical setup in S.I.~V. Under a pulsed two-photon resonant excitation (TPE) scheme\cite{Michler2014,Keil2017,Huber2017,Huber2018,Jayakumar2014}, the intensities of the XX and X recombination are comparable, since TPE populates the XX state, which feeds the X state. The cavity mode, with a quality factor of $\approx$150, is clearly identified via the reflectivity measurement (see more details in S.I.~VI) and it is resonant with both X and XX. The Purcell enhancement of the radiative decay of each state enabled by the cavity mode is directly quantified from time-resolved measurements in Fig.~2(b), showing comparisons of the lifetimes of X and XX in the CBR-HBR and in bulk (a different reference QD). We note the lifetimes of our reference QD in bulk are very consistent with the values reported in the similar systems\cite{Keil2017,Huber2017}. The lifetime of X is shortened from 210~ps to 60~ps by implementing the CBR-HBR, corresponding to a Purcell factor ($F_{p}$) of 3.5. A slightly higher Purcell factor of 4.4 for XX is obtained, due to a better spectral match to the cavity mode, which enables faster triggering rates of entangled photon pair emission compared to those of the QDs in bulk.

Second-order auto-correlation measurements are performed for both X and XX, see Fig.~2(c). The nearly complete absence of coincidence events at zero time delay indicates the ultra-high purity of the emitted single-photons. $g^{(2)}_{X}(0)=0.001(1)$ and $g^{(2)}_{X}(0)=0.007(1)$ are obtained without background correction. The slightly higher $g^{(2)}_{XX}(0)$ value of XX is mostly due to the very weak emission from neighboring QD states.

Different from the single-photon Rabi oscillation in which the population of the excited state is dependent on the pulse area, the pulse area in the two-photon Rabi oscillation is replaced by an adiabatic dynamic phase that turns out to be a non-trivial function of the quantum dot binding energy, pulse area, pulse duration and pulse shape\cite{Stufler2006}. Fig.~2(d) shows the detected photon flux from X as a function of the square root of the time-averaged excitation power. The Rabi oscillations of X and XX (not shown) are observed due to the coherent control of the two-level system consisting of the biexciton and crystal-ground-state in the QD. For this device, the photon count rate reaches a maximum for an average laser power of 36~nW, which we denote as "$\pi$ pulse" condition, similar to recent reports\cite{Michler2014,Huber2017}. We note that the laser power needed to reach "$\pi$ pulse" for the QDs in a CBR-HBR is at least 200 times lower than that for QDs in the bulk and also in simple planar cavities. Such a reduction of "$\pi$ pulse" power is attributed to the cavity enhanced excitation\cite{Kaniber2009} and represents an advantage for filtering the excitation laser in the entanglement and indistinguishability measurements we shall present.

\begin{figure*}[ht!]
	\begin{center}
		\includegraphics[width=0.8\linewidth]{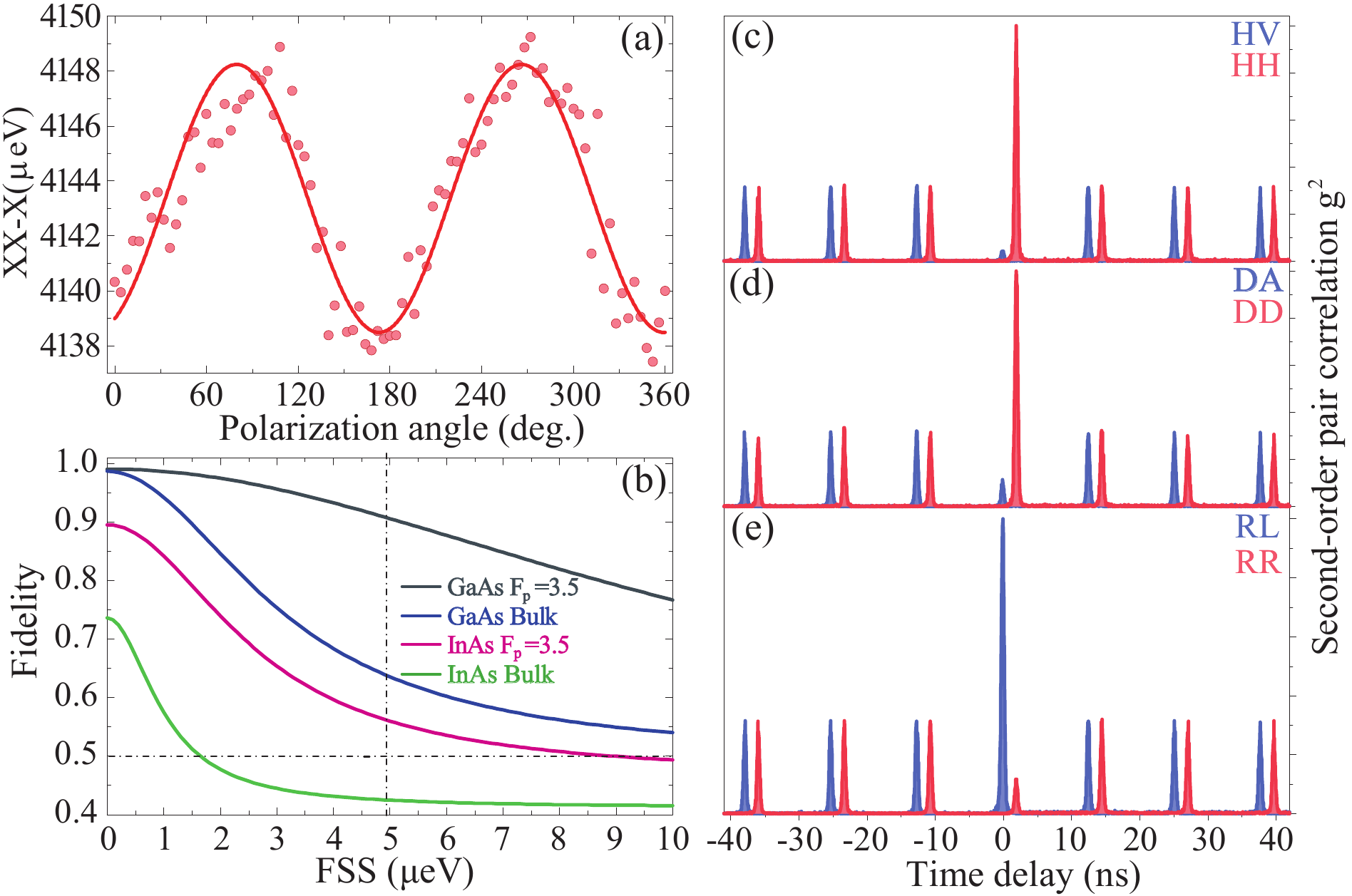}
		\caption{\textbf{Entanglement characterization.} Fidelity of the polarization entanglement is investigated. (a) Polarization-dependent measurement to determine the FSS of X. The relative energy difference between X and XX is plotted in order to obtain a higher measurement precision. An FSS value of 4.8(2)~$\mu$eV is extracted from the amplitude of the sine-function fitting. (b) Theoretically predicted entanglement fidelity as a function of FSS for GaAs QDs in the CBR-HBR ($F_{p}=3.5$, black line), in bulk (blue line), Purcell enhanced InAs QDs ($F_{p}=3.5$, ruby line) and InAs QDs in bulk (green line). The vertical dashed line denotes a FSS of 4.8~$\mu$eV and the horizontal dashed line ($f=0.5$) is the boundary above which quantum entanglement exists. (c), (d) and (e) are the X-XX polarization dependent cross-correlation histogram under "$\pi$ pulse" conditions for linear, diagonal, and circular basis respectively. Data for cross-polarization configurations are shifted deliberately for clarity.}
		\label{fig:Fig3}
	\end{center}
\end{figure*}

For a "$\pi$ pulse", we observe a photon count rate up to 3.4(1)~MHz under a 79~MHz repetition rate laser excitation. By taking the setup efficiency $\xi$ (7~\%, see S.I. VII ), avalanche photodiode (APD) correction factor ($1.25$), and XX preparation fidelity $\eta_{XX}$ ($\approx$0.9) into account, a collection efficiency (with a 0.65 NA objective) $\eta$=85(3)~\% is extracted for both X and XX. Consequently, the collected photon pair probability per pulse  $p=\eta_{XX}\times\eta^{2}\times[1-g^{(2)}_{X}(0)]^{1/2}\times[1-g^{(2)}_{XX}(0)]^{1/2}=0.65(4)$ is obtained. \textit{This high photon pair rate per pulse $p\approx0.65$ outperforms any of the existing entangled photon sources reported in the literature.}

\section*{Entanglement characterization}
The states of the photon pairs emitted by QDs can be written as  $\ket{\psi}=1/\sqrt{2}(\ket{H_{X}H_{XX}}+e^{is\tau/\hbar}\ket{V_{XX}V_{X}})$\cite{Michler2014,Keil2017,Huber2017,Shields2008}, in which $\tau$ is the, statistically varying, decay time of the XX state relative to the decay time of the X state and s is the value of the FSS. In absence of other dephasing mechanisms, the deviation of the two-photon states from the Bell state $\ket{\psi^{+}}$ originates from the phase factor $s\tau/\hbar$, which has to be minimized in order to obtain high level of entanglement without resorting to inefficient time-filtering\cite{Ward2014} or spectra-filtering\cite{Akopian2006}. Since the X lifetime $\tau_{X}$ is as short as 60~ps in our device thanks to the Purcell effect (corresponding to a lifetime-limited linewidth of $\sim$11~$\mu$eV), we expect that the generation of photon pairs with high time-averaged fidelity is still possible for QDs with a finite FSS. Using polarization-dependent measurements, shown in Fig.~3(a), a FSS of 4.8(2) $\mu$eV for the QD in the CBR-HBR is extracted by subtracting the X transition from the XX transition energy. The theoretical values of fidelity\cite{Hudson2007} as a function of FSS for different QDs with varied lifetimes are plotted in Fig.~3(b) (see the details in S.I.~VIII). For GaAs QDs with a Purcell factor of 3.5, the entanglement fidelity decays slowly with the increase of the FSS. The predicted fidelity for the GaAs QDs with a FSS of 4.8~$\mu$eV is as high as 0.92 and it can still be above 0.75 for a FSS of 10~$\mu$eV. On the contrary, the entanglement fidelity for GaAs QDs in bulk (lifetime of 210~ps) decreases much more quickly with the increase of the FSS and shows a slightly lower value than the Purcell-enhanced source at FSS=0. With the same FSS of 4.8~$u$eV, the entanglement fidelity is only 0.64 for GaAs QDs in bulk. In order to compare the performance against a different material system, we also plot the entanglement fidelities of Purcell-enhanced InAs QDs ($F_{p}=3.5$) and InAs QDs in bulk (typical lifetime of 1000~ps). For the InAs QDs in bulk, the highest fidelity is $<$ 0.75 and the entanglement disappears once the FSS is larger than 1.6~$\mu$eV. Even with the same Purcell factor of 3.5, the entanglement fidelity of InAs QDs is still not reaching to the level of GaAs QDs in bulk. These results are based on the spin-scattering times provided in Ref.~\onlinecite{Keil2017,Huber2017} and need further experimental confirmations. 

To evaluate the degree of entanglement of our bright photon-pair source, we perform cross-correlation measurements under $\pi$ pulse excitation for both X and XX photons in linear (HV), diagonal (DA), and circular (LR) basis sets. The cross-correlation histograms in the three basis sets are presented in Fig.~3(c-e). In linear and diagonal basis sets, we clearly observe the antibunching when the photon pairs are co-polarized and bunching for the cross-polarized photon pairs. The correlation in the circular basis is just opposite: co-polarized photon pairs show bunching while cross-polarized ones exhibit antibunching behavior. This set of correlations serves as a strong indication of polarization entanglement in the photon pairs. The degree of correlation in a particular polarization basis is defined by\cite{Hudson2007}:

\begin{equation*}
C_{\mu}=\frac{g^{(2)}_{XX,X}(0)-g^{(2)}_{XX,\bar{X}}(0)}{g^{(2)}_{XX,X}(0)+g^{(2)}_{XX,\bar{X}}(0)}
\end{equation*}


\noindent where $g^{(2)}_{XX,X}(0)$ and $g^{(2)}_{XX,\bar{X}}(0)$ are the second order correlation for the co-polarized and cross-polarized photon pairs in that basis at zero delay time. The extracted degree of correlation in the different basis sets from measured coincidence histograms are:
\begin{align*}
 C_{linear}=0.92(2) \\
 C_{diagonal}=0.81(2) \\
 C_{circular}=-0.80(2) \\
\end{align*}
With these numbers, the entanglement fidelity for the polarization between the emitted photon pairs can be calculated as:
\begin{equation*}
f=\frac{1+C_{linear}+C_{diagonal}-C_{circular}}{4}=0.88(2)
\end{equation*}
The theoretically predicted entanglement fidelity of 0.92 can be reduced to 0.88 by using a much shorter spin scattering time (1 ns instead of 15 ns), which strongly indicates the existence of extra dephasing processes. Such extra dephasing processes have been also recently observed in a similar material system and is attributed to the interactions between the confined exciton and charge states\cite{Huber2018}. In contrast to existing entangled sources with vanishing FSS, the pronounced Purcell effect in our work makes the high fidelity of entanglement possible for QDs with a comparatively large FSS. A near-unity entanglement fidelity can be expected in the future by implementing GaAs QDs with very small FSS in CBR-HBRs or by eliminating the residual FSS with a strain-tunable CBR-HBR\cite{Huber2018}(see strain-tunable CBR-HBR in the S.I.~IX).

\begin{figure*}
\begin{center}
\includegraphics[width=0.7\linewidth]{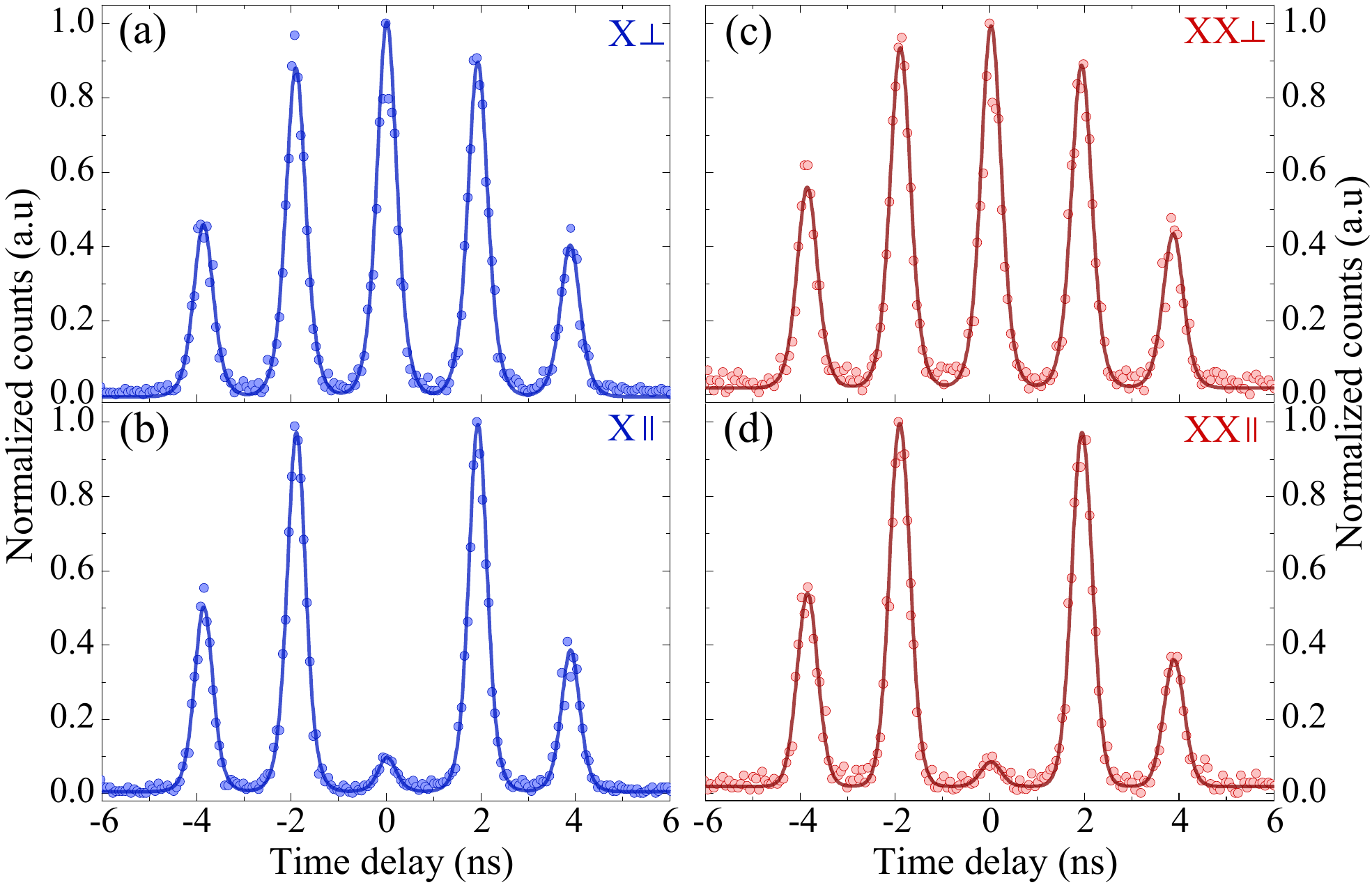}
\caption{\textbf{Photon indistinguishability.} HOM interference for X and XX photons are performed individually. Two-photon interference for cross-polarized (a), co-polarized (b) X photons and cross-polarized (c), co-polarized (d) XX photons. The data are fitted by exponential decays (measured emitter decay response) convolved with a Gaussian (measured photon detector time response). The area of the central peaks is extracted to calculate the raw visibilities, which are 0.901(3) and 0.903(3) for X and XX respectively.}
\label{fig:Fig4}
\end{center}
\end{figure*}

\begin{table*}[ht!]
	\centering
	\vspace{0.1cm}
	\begin{threeparttable}
		\caption{Characterization of more devices from the same chip.}
		\label{Table1}
		\begin{tabular}{|c|p{2.5cm}|c|c|c|c|c|c|}
			\hline
			\textbf{Device $\#$} \vspace*{\fill}  &   \textbf{Single-photon efficiency/Pair rate}  \vspace*{\fill} &    \textbf{Cavity wavelength (nm)} &    \textbf{X wavelength (nm)} & \textbf{X Purcell factor}  & \textbf{FSS ($\mu$eV)/f} & \textbf{X HOM} & \textbf{XX HOM}\\
			\hhline{|-|-|-|-|-|-|-|-|}
			\textbf{1} & 0.86/0.65 & 772.43 & 770.05 & 3.5 & 4.8/0.88 & 0.9 & 0.9 \\
			\hline
			\textbf{2} & 0.80 & 769.36 & 770.86 & 2.6 & 11.6/N.A. & N.A. & N.A. \\
			\hline
			\textbf{3} & 0.76 & 767.39 & 777.51 & 3.1 & 18.0/N.A. & N.A. & N.A. \\
			\hline
			\textbf{4} & 0.70/0.44 & 767.18 & 778.91 & 3.1 & 3.4/0.85 & 0.81 & 0.84 \\
			\hline
			\textbf{5} & 0.67 & 766.17 & 767.08 & 1.7 & 5.8/N.A. & N.A. & N.A. \\
			\hline
			\textbf{6} & 0.66 & 763.86 & 765.08 & 2.1 & 11.0/N.A. & N.A. & N.A. \\
			\hline
			\textbf{7} & 0.63 & 766.76 & 766.18 & 2.7 & 9.1/N.A. & N.A. & N.A. \\
			\hline
			\textbf{8} & 0.61 & 767.20 & 770.20 & 3.4 & 5.1/N.A. & N.A. & N.A. \\
			\hline
			\textbf{9} & 0.56 & 765.53 & 769.51 & 2.2 & 15.8/N.A. & N.A. & N.A. \\
			\hline
			\textbf{10} & 0.55 & 764.24 & 766.62 & 2.0 & 6.6/N.A. & N.A. & N.A. \\
			\hline
		\end{tabular}
	\end{threeparttable}
\end{table*}

\section*{Photon indistinguishability}
Photon indistinguishablity is a prerequisite for the realization of long-haul quantum information processing, e.g., a quantum repeater via entanglement swapping\cite{Duan2001}. We study the indistinguishablity of the emitted photons from our device via Hong-Ou-Mandel (HOM) interference measurements\cite{Santori2002}. The QD is excited  by two $\pi$-pulses separated by 1.9 ns with a repetition rate of 79~MHz (13~ns period). The emitted single photons are spectrally filtered within windows of ~$\sim$100~$\mu$eV, much larger than the zero-phonon-linewidths of the transitions, and projected to the horizontal polarization before being coupled to an unbalanced Mach-Zehnder interferometer (MZI) equipped with a 1.9~ns delay. A half-wave plate is placed in one arm of the MZI to prepare co-polarized or cross-polarized photons, making them distinguishable or indistinguishable in polarization. The emitted photons are interfered at the beam splitter in co- and cross-polarized configurations. The coincidence histogram of HOM interference for both X and XX are shown in Fig.~4. For both X and XX, the coincidence events at zero delay are greatly suppressed in the co-polarized configuration (Fig.~4(a,b)), which indicates the occurrence of two-photon interference at the beam splitter. In contrast, the area of the central peaks are almost the same as the side peaks in the cross-polarized configuration, showing no indistinguishability (Fig.~4(c,d)). Visibilities of two-photon interference of 0.901(3) and 0.903(3) for X and XX are extracted from the areas of the central peaks in the correlation histogram. We note that the high degree of photon indistinguishability in our pair source is a direct result of the Purcell effect, which has been recently shown as a key element to realizing highly indistinguishable single-photons from InAs QDs in micropillar cavities\cite{Ding2016,Somaschi2016,He2017}. With a joint force of further improvement of the Purcell factor, charge-stabilization\cite{Somaschi2016,Liu2018} and rapid adiabatic passage excitation\cite{Kaldewey2017}, higher photon indistinguishability can be expected as well as the entanglement fidelity. However, in our devices, the collection effiency tends to decrease with higher Q-factors (therefore higher Purcell factors) of the cavity due to the reduction of the overlap between the far-field pattern and the objective. Ultimately, the simultaneous realization of high collection efficiency and high indistinguishability will be fundamentally limited by the phonon scattering process. An upper bound can be placed on the indistinguishability by considering the microscopic theory developed in Ref.~\onlinecite{Smith2017}. This theory uses the polaron master equation formalism to capture non-Markovian phonon processes that lead to the emergence of a phonon sideband in the QD emission spectrum, and consequently degrades the indistinguishability of the source. Using this formalism with standard GaAs parameters, we find that our source could have an indistinguishability as high as 0.98 in the absence of any other dephasing processes (e.g. charge noise), see the S.I.~X for details of the calculation. However, moderate filtering of the phonon sidebands (at the expense of a few percent count reduction) may be used to further boost the indistinguishability.
\begin{table*}[ht!]
	\centering
	\vspace{0.1cm}
	\begin{threeparttable}
		\caption{Comparison of the performance of our device to the state-of-the-art entangled sources.}
		\label{Table1}
		\begin{tabular}{|c||c|c|c|}
			\hline
			\vspace*{\fill}  &   \textbf{Pair efficiency\tnote{a}}  \vspace*{\fill} &    \textbf{Entanglement Fidelity} &    \textbf{Indistinguishability} \\
			\hhline{|=||=|=|=|}
			\textbf{InAs QD in micropillar molecule (Ref~\onlinecite{Dousse2010})}        & 0.12 & 0.63  & Not shown \\
			\hline
			\textbf{InAsP QDs in nanowires (Ref~\onlinecite{Jons2017})}   & 0.0025   & 0.817 & Not shown\\
			\hline
			\textbf{InAs QDs in planar cavities (Ref~\onlinecite{Michler2014})}       & $<$0.0001   & 0.81 & 0.86 \\
			\hline
			\textbf{GaAs QDs in planar cavities (Ref~\onlinecite{Huber2017})}     & $<$0.0001   & 0.94  & 0.93 \\
			\hline
			\textbf{SPDC USTC (Ref~\onlinecite{Wang2016})}      & $\sim$0.1   & 0.93 & 0.91 \\
			\hline
			\textbf{SPDC Vienna (Ref~\onlinecite{Giustina2015})}       & $<0.01$   & $>$0.95 & $>$0.9 \\
			\hline
			\textbf{SPDC Geneva (Ref~\onlinecite{Geneva2005})}       & $<0.1$   & $\sim$0.9 & $\sim$0.9 \\
			\hline
			\textbf{This work}       & 0.65(4)   & 0.88(2) & $\sim$0.9 \\
			\hline
		\end{tabular}
		\begin{tablenotes}
			\item[a] The photon pair source efficiency is defined by the probability of collecting a photon pair per excitation pulse into the first collection optics, such as an objective or an optical fiber.
		\end{tablenotes}
	\end{threeparttable}
\end{table*}

While we focus on the performance of a single exemplary device, we have characterized several other devices on the same chip. In table~1 we have listed 10 such devices in the order of the brightness, among which 2 devices with the smallest FSSs are fully characterized (see S.I.~XI) and the others are partially characterized. The device 4 has smallest FSS of 3.4~$\mu$eV and the device 1 has the second smallest FSS of 4.8~$\mu$eV with the highest Purcell factor of 3.5. Therefore we do not expect any higher entanglement fidelity in the other devices. The limiting factor of the entanglement fidelity in this batch of devices is the relative large FSSs of the GaAs droplet QDs grown in the thin membrane (140~nm) structure with a thick sacrificial layer (500~nm), as shown in Fig.~S1(a). We believe such a limitation can be soon overcome by either optimizing the epitaxial growth process or developing the strain-tunable CBR-HBR that we proposed.

\section*{Summary}
Given the rapid development of the entangled photon sources both with SPDC and QD technologies, it is very insightful to directly compare the performance of our device to those of the existing sources reported in the literature. Table~2 lists the efficiency, entanglement fidelity, and indistinguishability of the state-of-the-art entangled photon sources together with our device (reference S.I.~XII for methodology used in extracting the various parameters.). In general, the SPDC sources exhibit excellent performance in terms of entanglement fidelity and photon indistinguishability; however, their efficiencies are intrinsically limited to $<$0.1 due to the nature of the Poissonian statistics. Increasing the photon pair flux through higher excitation power inevitably adds extra noise and reduces the purity and indistinguishability. For the deterministic approach, the efficiency of QDs in bulk suffers greatly from the total internal reflection and only a few works show a high-degree of indistinguishability. InAsP QDs in nanowires and InAs QDs in micropillar molecules show much improved brightness and decent entanglement fidelity, but still the source efficiency and indistingshability have to be further improved. Our device, for the first time, simultaneously combines a high pair collection probability ({0.65(4)}), high degree of entanglement fidelity (0.88(2)) and photon indistinguishability (0.901(3) and 0.903(3)), and when taken together outperforms all the existing entangled photon pair sources.

To conclude, we have implemented a broadband photonic nanostrucuture, CBR-HBR, to harvest highly-entangled photon pairs emitted by GaAs QDs, obtained by droplet etching. By employing the QD positioning technique based on fluorescence imaging, the QDs are accurately placed in the center of the CBR-HBR, thus enabling the realization of entangled sources with record performances. Our devices may immediately find applications in both fundamental physics and applied quantum technologies, e.g., quantum random walk with entangled photon pairs\cite{Pathak2007}, generation of hyper-entanglement\cite{Predojevic2017} and quantum repeaters\cite{Simon2007} associated with quantum memories. Moving forward, realizing high-performance photon pair sources operating in the telecom band\cite{Olbricha2017,Huwer2017} is particularly appealing for long-haul quantum communication. Instead of polarization entanglement, time-bin entanglements\cite{Jayakumar2014} can be directly generated from QDs, which makes our devices compatible with the fiber network. The operation wavelength for both droplet QDs and photonic nanostructures can be shifted to the telecom band by changing the filling material of the nanohole and scaling the size of the nanostructures. To scale this technology up to multiple QDs, piezo-tuning\cite{Yan2016,Trotta2016} or on-chip quantum frequency conversion technologies\cite{Li2016} can be directly implemented in our devices to tune the QD emission wavelength, overcome the spectral inhomogeneity between different QDs and eliminate the FSS. Such identical entangled pair sources can serve as individual nodes interconnected via single-photon interference in the future quantum network\cite{Gao2013}. With the potential of scalability, our work serves as a landmark in the development of semiconductor quantum information processing chips and may boost new breakthroughs in quantum photonic technologies.


\vspace{5mm} 
\noindent \textbf{Acknowledgements}\\
We acknowledge R.~Trotta, X.~Yuan, H.~Huang, M.~Reindl, D.~Huber and Y.~Huo for very fruitful discussions. We are greatful to the financial supports from  the National Key R\&D Program of China (2016YFA0301300, 2018YFA0306100), the National Natural Science Foundations of China (91750207, 1133405, 11674402, 11761141015, 11761131001, 11874437), Guangzhou Science and Technology project (201607020023, 201805010004), the Natural Science Foundations of Guangdong (2018B030311027, 2016A030312012), and the Austrian Science Fund (FWF): P29603.  
\vspace{5mm} 

\noindent \textbf{Author Contributions}\\
R.~B.~S, J.~T.~L and X.~H.~W conceived the nanostructure and its fabrication strategy. J.~L designed the measurement scheme. R.~B.~S and K.~S contributed to the structure simulations. S.~F.~CdS and Y.~Y grew the quatum dot wafers. R.~B.~S, B.~M.~Y, J.~T.~L and J.~L fabricated the devices. Y.~M.~W, R.~B.~S, B.~M.~Y and J.~L characterized the devices. J.~I.~S performed the indistinguishability calculation. J.~L, Y.~M.~W, and R.~B.~S analyzed the data. J.~L wrote the manuscript with inputs from all authors. J.~L, A.~R and X.~H.~W supervised the project.
\vspace{5mm} 

\noindent \textbf{Competing financial interests}\\
The author declare that they have no competing financial interests.
\vspace{5mm} 

\noindent \textbf{Data availability statement}\\
The data that support the plots within this paper and other findings of this study are available from the corresponding author upon reasonable request.
\vspace{5mm} 

\noindent \textbf{Additional information}\\
\textbf{Supplementary information} is available in the online version of the paper.\\ \textbf{Reprints and permission information} is available online at www.nature.com/reprints.\\
\textbf{Correspondence and requests for materials} should be addressed to A.~S, J.~T,~L or X.~H.~W.

\vspace{5mm} 
\noindent \textbf{Methods}\\
\noindent \textbf{Simulation}\\
The numerical simulaitons are carried out by means of finite-difference time-domain method, using a commercial software, Lumerical FDTD solutions. An electrical dipole is placed in the center of the CBR-HBR structure and six power monitors emcompassing the structure are employed to record power emitted by the dipole source. The sum power transmission normalized to that of the same source in homogeneous materials is calculated as the Purcell factor. The electric field record by the top monitor is used to calculate the far-field pattern by means of near-field to far-field projection. The collection efficiencies are extracted from the far-field distribution in 40 degree corresponding to N.A. of 0.65 (for further details see S.I. section II).

\noindent \textbf{Fabrication}\\
The epi-structure of the wafer is schematically shown in S.I.~I. After cleaning with acetone and isopropanol,  220~nm SiO$_{2}$ layer and 100~nm Au layer are deposited on the wafer by inductively coupled plasma chemical vapor deposition (Oxford instruments, PlasmaPro System100 ICP180-CVD) and electronic beam evaporation (Wavetest, DE400) repectively. The wafer is bonded to a glass substrate via ultraviolet curing resist (Norland, NOA 61). After exposure, the wafer is placed in a 50 $^{\circ}$C thermostat for 24 hours aging process to get an optimized performance. The original GaAs substrate is removed with phosphoric acid (H$\_{2}$PO$_{4}$:H$_{2}$O$_{2}$:H$_{2}$O = 1:1:1 volume) for 1.5 hour and selective etching solution (citric acid:H$_{2}$O$_{2}$ = 3:1 volume) until stopping at the sacrificial layer. The Al$_{0.8}$Ga$_{0.2}$As sacrificial layer is removed with 10\% HF. The CBR-HBR structures are defined by an electron beam lithography(Vistec EBPG5000+ system). The alignment marks (10~nm Ti and 100~nm Au) are patterned with an electron beam lithography and a lift-off process.  An Ar-SiCl$_{4}$ based dry etching process(Oxford instruments, PlasmaPro System 100 ICP180) is used to etch the GaAs structure.

\noindent \textbf{Two-photon resonant excitation}\\
A Ti-sapphire pulsed laser with a pulse duration of 120~fs and 79~MHz repetition rate is used to excited the QDs. In order to realize the two-photon resonant excitation scheme, the Ti-saphire laser is shaped by a home-made 4f-pulse shaper into a 8~ps pulse and spectrally tuned in the middle of the X and XX lines. The sample is excited by the optical pulses via an objective with a NA=0.65 and the emitted photon pairs are collected with the same objective. A notch filter is used to suppress the scattered laser background.


\section*{References}

\begin{thebibliography}{10}
	\bibitem{EPR}
	Einstein,~A., Podolsky,~B., and Rosen,~N.,
	\newblock Can Quantum-Mechanical Description of Physical Reality Be Considered Complete?,
	\newblock {\em Phys. Rev.} \textbf{47}, 777 (1935).
	
	
	\bibitem{Giustina2015}
	Giustina,~M., et al.
	\newblock Significant-Loophole-Free Test of Bell's Theorem with Entangled Photons.
	\newblock {\em Phys. Rev. Lett.} \textbf{115}, 250401 (2015).
	
	\bibitem{Shalm2015}
	Shalm,~L.~K., et al.
	\newblock Strong Loophole-Free Test of Local Realism.
	\newblock {\em Phys. Rev. Lett.} \textbf{115}, 250402 (2015).
	
	\bibitem{Bouwmeester_book}
	Bouwmeester,~D., Ekert,~A.~K., and Zeilinger,~A.
	\newblock {\em The Physics of Quantum Information} (Springer, 2000).
	
	\bibitem{Kimble2008}
	Kimble,~H.~J.
	\newblock The Quantum Internet.
	\newblock {\em Nature} \textbf{453}, 1023 (2008).
	
	\bibitem{Simon2007}
     Simon,~C., et al.
	\newblock Quantum Repeaters with Photon Pair Sources and Multimode Memories.
	\newblock {\em Phys. Rev. Lett.} \textbf{98}, 190503 (2007).

	\bibitem{Acin2007}
     Acin,~A., et al.
	\newblock Device-Independent Security of Quantum Cryptography against Collective Attacks.
	\newblock {\em Phys. Rev. Lett.} \textbf{98}, 230501 (2007).
	
	\bibitem{Kwiat1995}
     Kwiat,~P.~G., et al.
	\newblock New High-Intensity Source of Polarization-Entangled Photon Pairs.
	\newblock {\em Phys. Rev. Lett.} \textbf{75}, 4337 (1995).
	
	\bibitem{Geneva2005}
	Scarani,~V., Riedmatten,~H.~De, Marcikic,~I., Zbinden,~H., and Gisin,~N.
	\newblock Four-photon correction in two-photon bell experiments.
	\newblock {\em The European Physical Journal D-Atomic, Molecular, Optical and Plasma Physics} \textbf{32}, 129 (2005).
	
	\bibitem{Wang2016}
	Wang,~X.~L., et al.
	\newblock Experimental Ten-Photon Entanglement.
	\newblock {\em Phys. Rev. Lett.} \textbf{117}, 210502 (2016).
	
	\bibitem{Pan2012}
	Pan,~J.~W., et al.
	\newblock Multiphoton entanglement and interferometry.
	\newblock {\em Rev. Mod. Phys.} \textbf{2012}, 072501 (2012).

	\bibitem{Benson2000}
     Benson,~O., Santori,~C., Pelton,~M., and Yamamoto,~Y.
	\newblock Regulated and Entangled Photons from a Single Quantum Dot.
	\newblock {\em Phys. Rev. Lett.} \textbf{84}, 2513 (2000).

	\bibitem{Stevenson2006}
     Young,~R.~J., et al.
	\newblock A semiconductor source of triggered entangled photon pairs.
	\newblock {\em New J. Phys.} \textbf{8}, 29 (2006).

	\bibitem{Akopian2006}
     Akopian,~N., et al.
	\newblock Entangled Photon Pairs from Semiconductor Quantum Dots.
	\newblock {\em Phys. Rev. Lett.} \textbf{96}, 130501 (2006).

	\bibitem{Muller2009}
	Muller,~A., Fang,~W., Lawall,~J. and Solomon,~G~S.
     \newblock  Creating polarization-entangled photon pairs from a semiconductor quantum dot using the optical Stark effect.
	\newblock {\em Phys. Rev. Lett.} \textbf{103}, 217402 (2009).

	\bibitem{Michler2014}
	M\"uller,~M., Bounouar,~S., J\"ons,~K.~D., Gl\"ass,~M. and Michler,~P.
	\newblock On-demand generation of indistinguishable polarization-entangled photon pairs.
	\newblock {\em Nat. Photon.} \textbf{8}, 224 (2014).	

     \bibitem{Pelucchi2016}
	Chung,~T.~H., et al.
	\newblock Selective carrier injection into patterned arrays of pyramidal quantum dots for entangled photon light-emitting diodes.
	\newblock {\em Nat. Photon.} \textbf{10}, 782 (2016).	

     \bibitem{Orieux2017}
     Orieux,~A., Versteegh,~M.~A.~M., J\"ons,~K.~D. and Ducci,~S.
     \newblock Semiconductor devices for entangled photon pair generation: a review.
	\newblock {\em  Rep. Prog. Phys.} \textbf{80}, 076001(2017).

 	\bibitem{Huo2013}
	Huo,~Y.~H., Rastelli,~A. and Schmidt,~O~.~G.
	\newblock Ultra-small excitonic fine structure splitting in highly symmetric quantum dots on GaAs (001) substrate.
	\newblock {\em Appl. Phys. Lett.} \textbf{102},152105 (2013).

	\bibitem{Keil2017}
	Keil,~R., et al.
     \newblock Solid-state ensemble of highly entangled photon sources at rubidium atomic transitions.
	\newblock {\em Nat. Commun.} \textbf{10}, 15501 (2017).

     \bibitem{Huber2017}
	Huber,~D., et al.
     \newblock Highly indistinguishable and strongly entangled photons from symmetric GaAs quantum dots.
	\newblock {\em Nat. Commun.} \textbf{10}, 15506 (2017).
	
	
	\bibitem{Ding2016}
	Ding,~X., et al.
	\newblock On-demand single photons with high extraction efficiency and near-unity indistinguishability from a resonantly driven quantum dot in a micropillar.
	\newblock {\em Phys. Rev. Lett.} \textbf{116}, 020401 (2016).
	
	\bibitem{Somaschi2016}
	Somaschi,~N., et al.
	\newblock Near-optimal single-photon sources in the solid state.
	\newblock {\em Nat. Photon.} \textbf{10}, 340-345 (2016).
	
	\bibitem{He2017}
	He,~Y.-M., et al.
	\newblock Deterministic implementation of a bright, on-demand single-photon source with near-unity indistinguishability via quantum dot imaging.
	\newblock {\em Optica} \textbf{4}, 802-808 (2017).

     \bibitem{Claudon2010}
	Claudon,~J., et al.
	\newblock A highly efficient single-photon source based on a quantum dot in a photonic nanowire.
	\newblock {\em Nat. Photon.} \textbf{4}, 174 (2010)
	
	\bibitem{Reimer2012}
	Reimer,~M.~E., et al.
	\newblock Bright single-photon sources in bottom-up tailored nanowires.
	\newblock {\em Nat. Commun.} \textbf{3}, 737 (2012)

    \bibitem{Laucht2012}
    Laucht,~A., et al.
    \newblock A Waveguide-coupled On-Chip Single-Photon Source.
    \newblock {\em Phys. Rev. X} \textbf{2}, 011014 (2012).

    \bibitem{Arcari2014}
	Arcari,~M., et al.
	\newblock Near-unity coupling efficiency of a quantum emitter to a photonic crystal waveguide.
	\newblock {\em Phys. Rev. Lett.} \textbf{113}, 093603 (2014).

	\bibitem{Gschrey2015}
	Gschrey,~M., et al.
	\newblock Highly indistinguishable photons from deterministic quantum-dot microlenses utilizing three-dimensional in situ electron-beam lithography.
	\newblock {\em Nat. Commun.} \textbf{6}, 7662 (2015).

	\bibitem{Davanço2011}
	Davanco,~M., Rakher,~M.~T., Schuh,~D., Badolato,~A. and Srinivasan,~K.,
	\newblock A circular dielectric grating for vertical extraction of single quantum dot emission.
	\newblock {\em Appl. Phys. Lett.} \textbf{99}, 041102 (2011).
	
	\bibitem{Sapienza2015}
	Sapienza,~L., Davanço,~M., Badolato,~A. and Srinivasan,~K.
	\newblock Nanoscale optical positioning of single quantum dots for bright and pure single-photon emission.
	\newblock {\em Nat. Commun.} \textbf{6}, 7833 (2015).


	\bibitem{Dousse2010}
      Dousse,~A., et al.
	\newblock Ultrabright source of entangled photon pairs.
	\newblock {\em Nature} \textbf{466}, 217 (2010).

	\bibitem{Jons2017}
     J\"ons,~K.~D., et al.
	\newblock Bright nanoscale source of deterministic entangled photon pairs violating Bell’s inequality.
	\newblock {\em Sci. Rep.} \textbf{7}, 1700 (2017).
	
	\bibitem{Chen2018}
    Chen,~Y., Zopf,~M., Keil,~R., Ding,~F. and Schmidt,~O.~G.
	\newblock Highly-efficient extraction of entangled photons from quantum dots using a broadband optical antenna.
	\newblock {\em Nat. Commun.} \textbf{9}, 2994 (2018).
	

     \bibitem{Liu2017_RSI}
 	 Liu,~J., et al.
	 \newblock Cryogenic photoluminescence imaging system for nanoscale positioning of single quantum emitters.
	 \newblock {\em Rev. Sci. Instrum.} \textbf{88}, 023116 (2017).
	 
	 \bibitem{Yan2016}
	 Chen,~Y., et al.
	 \newblock Wavelength-tunable entangled photons from silicon-integrated III–V quantum dots.
	 \newblock {\em Nat. Commun.} \textbf{7}, 10387 (2016).
	 
	 \bibitem{Trotta2016}
	 Trotta,~R., et al.
	 \newblock Wavelength-tunable sources of entangled photons interfaced with atomic vapors.
	 \newblock {\em Nat. Commun.} \textbf{7}, 10375 (2016).
	 
	 \bibitem{Huber2018}
	 D.~Huber et al.
	 \newblock Strain-tunable GaAs quantum dot: An on-demand source of nearly-maximally entangled photon pairs.
	 \newblock {\em Phys. Rev. Lett.} 121, 033902 (2018).

	\bibitem{Jayakumar2014}
	Jayakumar,~H., et al.
	\newblock Time-bin entangled photons from a quantum dot.
	\newblock {\em Nat. Commun.} \textbf{5}, 4251 (2014).
	
	\bibitem{Stufler2006}
	Stufler,~S., et al.
	\newblock "Two-photon Rabi oscillations in a single $\mathrm{In_{x}Gas_{1-x}A/GaAs}$ quantum dot,"
	\newblock {\em Phys. Rev. B} \textbf{73}, 125304 (2006).

    \bibitem{Kaniber2009}
 	Kaniber,~M., et al.
	\newblock Efficient and selective cavity-resonant excitation for single photon generation.
	\newblock {\em New J. Phys.} \textbf{11}, 013031 (2009).

    \bibitem{Shields2008}
	Stevenson,~R.~M., et al.
	\newblock Evolution of Entanglement Between Distinguishable Light States.
	\newblock {\em Phys. Rev. Lett.} \textbf{101}, 170501 (2008).

	\bibitem{Ward2014}
	Ward,~M.~M., et al.
     \newblock Coherent dynamics of a telecom-wavelength entangled photon source.
	\newblock {\em Nat. Commun.} \textbf{5}, 3316 (2014).


    \bibitem{Hudson2007}
	Hudson,~A.~J., et al.
	\newblock Coherence of an Entangled Exciton-Photon State.
	\newblock {\em Phys. Rev. Lett.} \textbf{99}, 266802 (2007).


	\bibitem{Duan2001}
     Duan,~L.~M., Lukin,~M.~D., Cirac,~J.~I. and Zoller,~P.
	\newblock Long-distance quantum communication with atomic ensembles and linear optics.
	\newblock {\em Nature} \textbf{414}, 413 (2001).

     \bibitem{Santori2002}
     Santori,~C., Fattal,~D., Vuckovic,~J., Solomon,~G.~S. and Yamamoto,~Y.
	\newblock Indistinguishable photons from a single-photon device.
	\newblock {\em Nature} \textbf{419}, 594 (2002).
	
    \bibitem{Liu2018}
	Liu,~J., et al.,
	\newblock Single Self-Assembled InAs/GaAs Quantum Dots in Photonic Nanostructures: The Role of Nanofabrication.
	\newblock {\em Phys. Rev. Appl.} \textbf{9}, 064019 (2018).
	
	\bibitem{Kaldewey2017}
	Kaldewey,~T., et al.
	\newblock Coherent and robust high-fidelity generation of a biexciton in a quantum dot by rapid adiabatic passage.
	\newblock {\em Phys. Rev. B} \textbf{95}, 161302(R) (2017).
	
	\bibitem{Smith2017}
    Iles-Smith,~J., McCutcheon,~D.~P.~S., Nazir,~A., and Mork,~J.,
	\newblock Phonon scattering inhibits simultaneous near-unity efficiency and indistinguishability in semiconductor single-photon sources.
	\newblock {\em Nat. Photon.} \textbf{11}, 521 (2017).	

    \bibitem{Pathak2007}
	Pathak,~P.~K. and Agarwal,~G.~S.
	\newblock Quantum random walk of two photons in separable and entangled states.
	\newblock {\em Phys. Rev. A} \textbf{75}, 032351 (2007).

    \bibitem{Predojevic2017}
	Prilm\"uller,~M., et al.
	\newblock Hyperentanglement of Photons Emitted by a Quantum Dot.
	\newblock {\em Phys. Rev. Lett.} \textbf{121} 110503 (2018).

     \bibitem{Olbricha2017}
	Olbricha,~F., et al.
	\newblock Polarization-entangled photons from an InGaAs-based quantum dot emitting in the telecom C-band.
	\newblock{\em Appl. Phys. Lett.}, \textbf{111}, 133106 (2017).

    \bibitem{Huwer2017}
	Huwer,~J., et al.
	\newblock Quantum-Dot-Based Telecommunication-Wavelength Quantum Relay.
	\newblock {\em Phys. Rev. Appl.} \textbf{8}, 024007 (2017).

	\bibitem{Li2016}
	Li,~Q., Davan\c co,~M. and Srinivasan,~K.
	\newblock Efficient and low-noise single-photon-level frequency conversion interfaces using silicon nanophotonics.
	\newblock {\em Nat. Photon.} \textbf{10}, 406 (2016).

	\bibitem{Gao2013}
	Gao,~W.~B., et al.
	\newblock Quantum teleportation from a propagating photon to a solid-state spin qubit.
	\newblock {\em Nat. Commun.} \textbf{11}, 2744 (2013)。


	
	
	
\end{thebibliography}
\end{document}